# Revealing sub-μm inhomogeneities and μm-scale texture in $H_2O$ ice at Megabar pressures via sound velocity measurements by time-domain Brillouin scattering

Sergey M. Nikitin[1], Nikolay Chigarev[1], Vincent Tournat[1], Alain Bulou[2], Damien Gasteau[1], Bernard Castagnede[1], Andreas Zerr[3], and Vitalyi E. Gusev[1]

[1]LAUM, UMR-CNRS 6613, Université du Maine, Le Mans, France
[2]IMMM, UMR-CNRS 6283, Université du Maine, Le Mans, France
[3]LSPM, UPR-CNRS 3407, Université Paris Nord, Villetaneuse, France

**Abstract**

Time-domain Brillouin scattering technique, also known as picosecond ultrasonic interferometry, which provides opportunity to monitor propagation of nanometers to sub-micrometers length coherent acoustic pulses in the samples of sub-micrometers to tens of micrometers dimensions, was applied to depth-profiling of polycrystalline aggregate of ice compressed in a diamond anvil cell to Megabar pressures. The technique allowed examination of characteristic dimensions of elastic inhomogeneities and texturing of polycrystalline ice in the direction normal to the diamond anvil surfaces with sub-micrometer spatial resolution via time-resolved measurements of variations in the propagation velocity of the acoustic pulse traveling in the compressed sample. The achieved two-dimensional imaging of the polycrystalline ice aggregate in-depth and in one of the lateral directions indicates the feasibility of three-dimensional imaging and quantitative characterization of acoustical, optical and acousto-optical properties of transparent polycrystalline aggregates in diamond anvil cell with tens of nanometers in-depth resolution and lateral spatial resolution controlled by pump laser pulses focusing.

*Corresponding authors: vitali.goussev@univ-lemans.fr, zerr@univ-paris13.fr, sergey.nikitin.Etu@univ-lemans.fr

### 1. Introduction

Knowledge of pressure-dependences of sound velocities and elastic moduli of liquids and solids (amorphous or crystalline) at Megabar pressures and evolution of texture of polycrystalline solids on compression is of extreme importance for a few branches of natural sciences such as condensed matter physics, physics of the Earth and planetology, as well as for monitoring and predicting of earthquakes and tsunamis or nuclear weapons test control.[1,2] The high-pressure parameters of materials, which can be measured at pressures above 25 GPa in laboratory conditions only using a diamond anvil cell (DAC), can be importantly influenced by the polycrystallinity of the material and the existence of spatial inhomogeneities and texture in polycrystalline aggregates even in the samples of several μm characteristic size.[3] It is well known that the aggregate sound velocity depends on the characteristic dimensions of the individual crystallites, which are elastically anisotropic,[4] and that, even a partial alignment of the crystallites in a material, i.e., orientational texture, causes the modification of the measured aggregate sound velocities.[5] Both these factors are preventing precise evaluation of the elastic moduli on the basis of sound velocity measurements and make necessary the development of the experimental methods for the three-dimensional imaging of the microscopic samples *in-situ* when compressed in a high-pressure apparatus with the goal of characterization both their morphological and orientational/directional texture.

In the past decade significant progress has been achieved in X-ray imaging of isolated objects and multigrain bodies with the spatial resolution better than 100 nm in cases where inhomogeneities have a chemical contrast e.g. Ref[6]. In the cases where grain orientations in polycrystalline bodies are derived from X-ray diffraction, the spatial resolution approaches today submicron level.[7,8] These methods based on the use of synchrotron X-ray radiation have also been applied for examination of samples recovered from high-pressure experiments.[9,10] However, a limited access to the sample compressed in a diamond anvil cell reduces significantly the lateral resolution for *in-situ* characterisation of polycrystalline samples.[11,12] The resolution degradation in the axial sample direction, i.e., along in-depth direction parallel to the load axis of the DAC, is even more dramatic and was expected to be 2 μm or even worse.[13] The nanoscale resolution can be maintained for samples compressed in a DAC only when a small number of grains or an isolated grain are illuminated, or when the sample is chemically and texturally uniform (e.g. single crystal or glass).[14-16] Finally, none of the considered methods provides any independent information on elastic properties of the sample material or individual grains. In the present paper we introduce an experimental method which overcomes the above listed limitations and allows *in-situ* texture examination of transparent samples compressed in DAC to pressures approaching 1 Mbar and provides spatially resolved information (both in lateral and axial directions) about sample elastic behaviour.

Among the methods for the evaluation of the elastic parameters of the condensed media at high pressures important role is played by those based on the interaction of laser radiation with acoustical phonons. Application of optical radiation is particularly suited for the DAC technique where both extreme stiffness-hardness, needed to achieve high pressures, and transparency of diamond, needed for optical access to compressed sample, are essential. In classic, i.e., frequency domain, Brillouin scattering (BS) technique the frequency shift of light scattered by thermal (incoherent) acoustic phonons in the sample provides information on the velocities of the longitudinal and shear sound.[17-19] In laser ultrasonic techniques light is employed not only for the detection of acoustic phonons, like in classic BS, but also for their generation.[20,21] Moreover, the acoustic phonons, monitored inside the sample, are coherent. Laser ultrasonics had been first used in high pressure experiments to increase the photo-elastic scattering of probe light through the generation of monochromatic coherent acoustic waves by laser-induced gratings, i.e., by interference patterns of two pump pulsed laser beams propagating at an angle. The probe light is predominantly scattered not by the thermal but by the laser-generated phonons. This laser ultrasonic technique of impulsive stimulated Brillouin scattering (ISBS)[22] had been applied at high pressures to measure velocities of both bulk[23,24] and interface[25,26] acoustic waves. Recently other laser ultrasonic techniques were tested in studies of materials at high pressures in DACs. Picosecond laser ultrasonic technique,[27] based on both the generation and the detection of wide-frequency-band coherent acoustic pulses by femtosecond laser pulses was first applied in Ref.[28]. Laser ultrasonic technique based on the generation of acoustic pulses by sub-nanosecond laser pulses and their detection by the continuous laser radiation was introduced in Ref.[29]. In both experimental configurations the spectrum of the detected acoustic pulses extended up to GHz frequencies range. In Ref.[28] the transient optical reflectivity signals obtained by the technique of picosecond acoustic interferometry,[30] also called time-domain Brillouin scattering (TDBS), were reported for the first time at high pressures. Later this technique has been also realized with picosecond laser pulses instead of femtosecond laser pulses.[31]

In the picosecond acoustic interferometry, which is a particular optical pump – probe technique, the pump laser pulse generates in a light absorbing opto-acoustic transducer picosecond acoustic pulse which propagates through a sample. Since a typical length of picosecond acoustic pulse is in the nanometers to sub-micrometers spatial scale, the technique is perfectly suitable for examination of materials confined in DACs where sample sizes are typically from several tens of micrometers down to a few micrometers and grains in polycrystalline samples are typically below 1 μm in size (Fig. 1 (b)). In case of an optically transparent material the probe laser pulse, delayed in time relative to the pump laser pulse, preferentially interacts with those coherent GHz phonons of the acoustic pulse spectrum which satisfy the momentum conservation law in photon-phonon photo-elastic interaction, i.e., satisfy the BS condition. Weak light pulses scattered by acoustic pulse interfere at the photo-detector with the probe light pulses of significantly higher amplitude scattered/reflected at various surfaces and interfaces of the tested structure, for example at the interfaces of the sample with the diamond and of the opto-acoustic transducer with the sample (Fig.1 (b)). The detected modification of the transient optical reflectivity is proportional, in leading order, to the product of these two scattered light fields. Thus, a heterodyning of a weak field against a strong one is achieved in picosecond ultrasonic interferometry. The measured transient reflectivity signal varies with time because the relative phase of the light scattered by the propagating acoustic pulse and scattered by immobile surfaces/interfaces continuously changes with time due to the variation in the spatial position of the propagating acoustic pulse. If acoustic pulse propagates at a constant velocity, i.e., in a spatially homogeneous medium, the phase difference between the interfering light fields linearly changes in time and, as a consequence, the amplitude of the signal changes in time in sinusoidal manner at a GHz frequency precisely equal to the Brillouin frequency.[30] Thus, measuring the period/frequency of this *time-domain* Brillouin oscillation provides information on the velocity of the acoustic wave in the sample. In the collinear scattering geometry of the TDBS experiments[28,31] the Brillouin frequency is proportional to the product of sound velocity and of the optical refractive index of the medium at the probe wave length (see Eq. (1) in Results).

In an inhomogeneous medium the TDBS signal at each time instance contains information on the local parameters of the medium in the spatial position of the laser-generated light-scattering acoustic pulse at this time instance. It has been demonstrated earlier, although under ambient conditions, that this effect can be used for the depth-profiling of inhomogeneous transparent media with nanometers scale resolution limited by the spatial length of the laser-generated coherent acoustic pulse.[32,33] The in-depth profiles of the sound velocity, of the optical refractive index and of the photo-elasticity with the spatial resolution of 50 nm, which is not limited by the wavelength of acoustic phonon at Brillouin frequency, were determined in the sub-micrometer thick film of low-$k$ nanoporous material by conducting TDBS measurements at several different angles of probe light incidence.[32]

Below we report on application of the TDBS technique to the diagnostics of water ice in a DAC at pressures of 50, 57 and 84 GPa revealing the characteristic features of its micro-crystallinity. While the classic, i.e., frequency domain, BS experiments with single crystal $H_2O$ ice are extremely rare[34,35] and were reported only at relatively low pressures below 10 GPa, the experiments at higher pressures are conducted on polycrystalline ice.[34,36-38] The problems in application of the classic BS of light by thermal phonons in polycrystalline materials and, in particular, in $H_2O$ ice are well known.[19,36-38] The Brillouin spectral lines are not just importantly broadened[19,36] but are split[19,36,38] because of the simultaneous contributions to the scattered light from multiple differently oriented crystallites, i.e., elastically anisotropic grains, inside the scattering volume. Detection of the broad Brillouin lines provides opportunity to extract only the orientation-averaged, i.e., so-called aggregate, sound velocities,[36] while the detection of the split Brillouin peaks requires data collection through a large number of different scattering volumes covering many randomly oriented grains with the goal to establish the maximum and the minimum boundaries of compressional and shear sound velocities.[38] Additional complexity/inconvenience in the experiments with cubic polycrystalline ices arrives from their optical isotropy,[39] which makes impossible visualization/characterization of the grain distribution, i.e., of the polycrystalline aggregate texture, by birefringence, which could be very useful in the case of grains without inversion symmetry.[34,40] In-depth spatial resolution in classic BS microscopy[41] applied for three-dimensional imaging[42] exceeds nowadays tens of micrometers.

Here we report experimental "visualization" of the texture in the polycrystalline aggregate of $H_2O$ ice by TDBS technique. The demonstrated two-dimensional, in-depth and lateral, imaging of texture in cubic, i.e., optically isotropic, ice is mostly due to the contrast provided by the difference in sound wave velocities inside the crystallites, differently oriented relative to the propagation direction of the acoustic and probe laser pulses. When coherent acoustic pulse of the nanometers to sub-micrometers spatial length propagates inside the sample, we resolve in time/space the Brillouin frequencies corresponding to the particular orientations of the spatial domains, in which the coherent acoustic pulse is currently localized. The lateral dimension of the photo-generated acoustic pulse is controlled by pump laser pulse focusing.

## 2. Results

The experiments on samples compressed in a DAC were conducted using typical pump/probe configuration for transient reflectivity optical measurements presented in Fig. 1 (a) (see Methods).

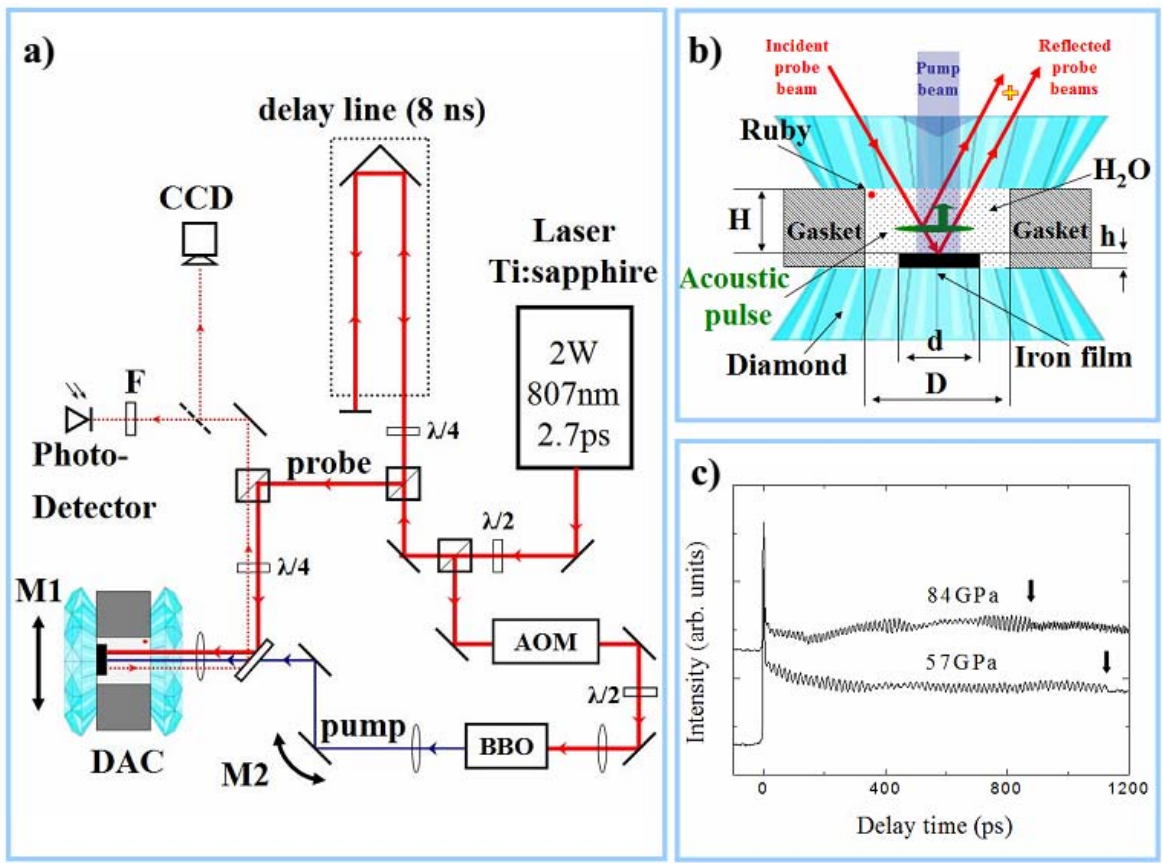

**Figure 1. (a):** Schematic presentation of the experimental setup. **(b):** Diamond anvil cell with qualitative presentation of some of the probe optical rays contributing to time-domain Brillouin scattering detection. For the characteristic dimensions of the ice sample and of the opto-acoustic transducer noted in the insert see Methods. **(c):** Transient reflectivity signals detected in the $H_2O$ ice samples compressed in a DAC to pressures of 57 and 84 GPa. Arrows indicate the delay times for the first arrival of the photo-generated acoustic pulse at the interface of ice and diamond.

### 2.1 Experimental signals and their processing

Typical transient reflectivity signals recorded in the experiments at 57 and 84 GPa where water crystallizes, most probably, in the phase ice X with ordered positions of protons[43,44] are presented in Fig. 1 (c). The temporal window includes the delay times necessary for a one-way propagation of the photo-generated coherent acoustic pulse from the iron opto-acoustic generator to the ice/diamond interface (see Fig. 1 (b)). The arrival times of the acoustic echoes at ice/diamond interface, marked by arrows in Fig. 1 (c), manifest themselves by an abrupt diminishing in the amplitude of the Brillouin oscillation at 880 ps and 1134 ps for 84 GPa and 57 GPa, respectively. In Fig. 1 (b) and especially in the insert in Fig. 2 (a) it is also clearly seen, without any signal processing, that the transmission of the acoustic pulse across ice/diamond interface is accompanied by an abrupt increase in the frequency of the oscillating part of transient reflectivity. These observations are in accordance with the estimated weak reflection, of less than 15 % in amplitude, of acoustic waves from ice/diamond interface, much lower photo-elastic (acousto-optic) constants of diamond in comparison with those of ice at the optical probe wavelength and larger Brillouin frequency shifts in diamond than in $H_2O$ ice at the considered pressures measured by classic BS.[36] The measurement of these arrival times is known to provide information on the acoustic velocity averaged over the path of the coherent acoustic pulse propagation and on the thickness of the sample.[32,33,45-47] In the case of a polycrystalline aggregate, the measurements of these arrival times provide information on the aggregate sound velocity (see Discussion). The observed abrupt changes in the amplitude and frequency of the Brillouin oscillations upon the arrival of the acoustic pulse on the ice/diamond interface indicate that the duration of the acoustic pulse does not exceed the Brillouin period (its spatial length does not exceed the acoustic wave length at Brillouin frequency).

The non-monotonous in time variations in the amplitude of the Brillouin oscillations, which are observable in Figs. 1 (c) and 2 without any signal processing, is a strong indication of the spatial inhomogeneity, i.e., micro-crystallinity and texturing, of the samples. They cannot be attributed to the beatings of TDBS signals from quasi-longitudinal and quasi-shear coherent acoustic pulses and are manifestations of texturing in the distribution of both photo-elastically (acousto-optically) and elastically anisotropic grains composing the micro-crystalline ice aggregate (see Section 2.2.)

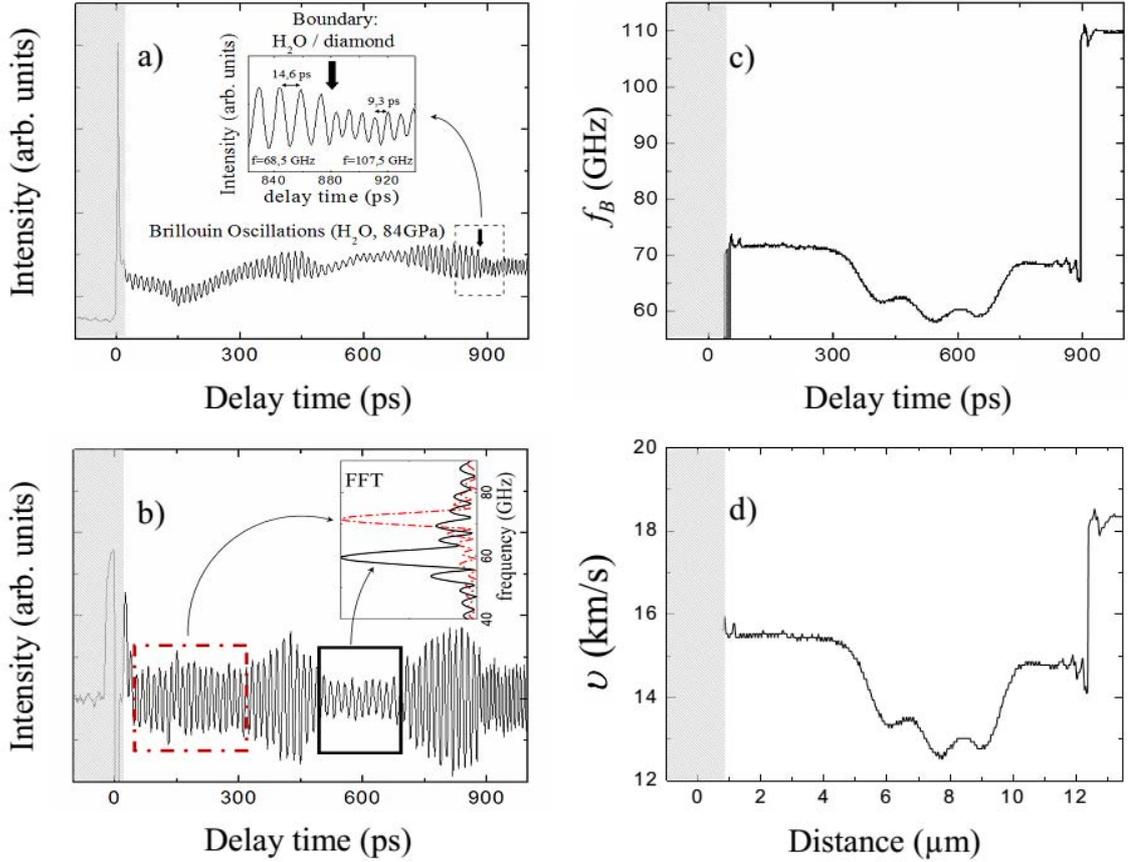

**Figure 2.** (a): Typical time-resolved reflectivity signal in $H_2O$ compressed in a DAC to 84 GPa. Vertical arrow marks the time of transmission of the laser-generated acoustic pulse across the interface of $H_2O$ ice X with diamond. Insert: zoom of the signal in the vicinity of the ice/diamond interface. (b) Time-domain Brillouin scattering signal obtained by filtering from the signal in (a) of the transient thermo-reflectance contribution caused by transient heating of the sample. Insert: Fourier spectra of the signal inside two temporal windows, marked by red rectangular and black rectangular, demonstrate increase of the Brillouin frequency from 60 GHz up to about 73 GHz, indicating spatial inhomogeneity of the $H_2O$ ice X sample. (c): Temporal profile of the Brillouin frequency of the TDBS signal shown in (b). Fourier transform was performed in the rectangular moving temporal window of 100 ps duration which corresponds to the in-depth distance of about 1.5 μm. (d): Spatial variation of the longitudinal sound velocity obtained from the temporal dependence of the backscattering sound velocity shown in (c) using the value of the refractive index of $H_2O$ ice extrapolated from the experimental data in Ref.[39].

Additional indication of the spatial inhomogeneity could be the variation with time of the frequency of Brillouin oscillation. In the bulk of ice X sample this can be revealed by Fourier transform performed inside of an appropriately chosen time window after filtering from the total time-resolved reflectivity signal of the time-varying contribution of thermo-reflectance caused by sample heating.[32,33] This signal processing reveals the time-domain Brillouin scattering (TDBS) signal which is due to the interaction of the probe light with photo-generated acoustic pulse only (Fig. 2 (b)). The insert in Fig. 2 (b) demonstrates that the Fourier transform of the signal inside two fixed temporally separated windows, marked by red and black rectangular, reveals the Brillouin frequencies of 73 GHz and 60 GHz, respectively. These 10% deviations of frequencies from their average clearly indicate the presence of spatial elastic inhomogeneities in the ice X sample due to micro-crystallinity/texturing. The reader should bear in mind here that for each time-window the obtained frequency value is an average over multiple crystallites in the volume from which the signal is collected. The lateral size of the probed volume is determined by the product of the probe beam intensity distribution and pump beam intensity distribution, to which the strain amplitude in coherent non-diffracting acoustic beam is proportional. For the pump and probe beams elliptical intensity correlation function with the axes of 3 µm and of 4.5 µm FWHM (see Methods) the cross sectional area of the tested volume can be estimated as 10 µm$^2$. The axial, i.e., the in-depth, dimensions of the probed sample volumes for the results presented in insert in Fig. 2 (b), are determined by the duration of the used time-windows and exceed, in this particular case, 3 µm.

Performing Fourier transform in a continuously moving smoothed time-window of appropriately chosen duration and taking the frequency of the maximum in the spectra, it is possible to extract the temporal-profile of the dominant Brillouin frequency along the particular path of the coherent acoustic pulse propagation with different degrees of averaging along the sound propagation direction. In depth-profiling of the spatially inhomogeneous media the in-depth spatial resolution of TDBS is known to be limited either by the spatial length of the coherent acoustic pulse or by restrictions introduced by specific signal processing techniques.[32,33] In our experiments the duration of the emitted longitudinal acoustic pulse is controlled by the pump laser pulse duration of 1.9 ps and its width in ice near the opto-acoustic generator is about 30 nm at 84 GPa (see Section 2.3). The experimental results in Fig. 2 (a) demonstrate that although this pulse could be broadened by bulk high-frequency absorption and scattering, its duration at the ice/diamond interface still does not exceed the Brillouin period of about 15 ps corresponding to about 0.2 µm spatial scale at 84 GPa. In the current report we will use for the Fourrier transforms the windows exceeding at FWHM the Brillouin periods. Thus we achieve for the demonstration purposes depth-profiling with the same homogeneous in-depth spatial resolution controlled by the duration of the moving window in the complete experimental time domain by the simplest signal processing technique. Achieving of in-depth spatial resolution controlled by the duration of the acoustic pulse by applying advanced methods of signal processing[32,33] is beyond the scope of the present report.

In Fig. 2 (c) we present the temporal profile of the Brillouin frequency for TDBS signal shown in Fig. 2 (b) obtained with the moving rectangular time-window of 100 ps corresponding to the in-depth spatial resolution of about 1.5 μm.

The frequency of the detected time-resolved Brillouin oscillation is controlled by the momentum conservation law in the photon scattering by the acoustic phonon.[30,32,45,47] In our polycrystalline ice sample the photo-generated coherent acoustic pulse, which is initially launched normally to the Fe/ice interface and propagates parallel to the direction of the probe laser beam, could be refracted when crossing such plane interfaces between the grains, which are not parallel to the Fe/ice interface. Thus, in general, the coherent acoustic pulse propagates non-collinearly to the probe light. In this case the evaluation of the triangle composed of the wave vectors of an acoustic phonon and of the incident and of the scattered probe light photons leads to the following solution for the Brillouin frequency $f_B$.[32]

$$f_B \cong 2n\upsilon\cos\theta / \lambda_0. \tag{1}$$

Here $\upsilon$ denotes the speed of the acoustic wave along particular direction of its propagation in elastically anisotropic media, $\lambda_0$ is the wavelength of the probe light in vacuum, $n$ is the optical refractive index of the media for the probe light and $\theta$ is the angle between the propagation directions of the probe light and the sound. Note that in deriving Eq. (1) it is taken into account that frequency of light is importantly higher than the frequency of sound (quasi-elastic scattering of light). In the general case, three factors in Eq. (1), i.e., $\theta$, $n$ and $\upsilon$, could be responsible for the different values of the Brillouin frequency, when the coherent acoustic pulse travels along its propagation path. We estimated theoretically and confirmed by additional experiments that variations of the refractive index are playing negligible role in comparison with those of sound velocity (see Section 2.4) In Section 2.5 we demonstrate theoretically that variations caused by the acoustic beam refractions could be also neglected in the first approximation. Thus in the following we assume that the experimentally revealed variations in the Brillouin frequency are due to the variations of the sound velocity only, i. e., that they are dominantly caused by the different orientations of elastically anisotropic grains or groups of grains in the polycrytstalline ice aggregate. As an example, assuming in Eq. (1) $\theta = 0$ and taking the refractive index of the ice from the literature[39] it is straightforward to obtain from the dependence in Fig. 2 (c) the dependence of the acoustic velocity in ice first on time and then on the in-depth coordinate (Fig. 2 (d)). Note, that the coordinate in Fig. 2 (d) is evaluated by integrating the sound velocity over acoustic propagation time only in a part of the experimental window, which does not include short delay times (shadowed in Fig. 2) where the determination of Brillouin frequency is not enough certain because of imperfect filtering of the thermo-reflectance contribution at GHz frequencies. Thus, the distance in Fig. 2 (d) should be measured relative to the ice/diamond interface. The depth profile of the longitudinal acoustic velocity in Fig. 2 (d) clearly demonstrates that even on the micrometer scale the relative variations of the acoustic velocity, when the coherent acoustic pulse propagates through our polycrystalline ice aggregate, can exceed ± 10%, thus indicating a strong elastic anisotropy of differently orientated crystallites or their groups.

Experimentally observed inhomogeneity of the polycrystalline aggregates of ice X can be deeper characterized in two-dimensional imaging experiments, performing the measurement in several consecutive neighbor spatial positions along the lateral surface of the sample. In Fig. 3, in order to reveal the large scale inhomogeneity of ice X aggregate, we present the images of the Brillouin frequency distribution, which were obtained with the spatial resolution of about 0.9 μm, determined by the half-width at half-maximum of the Hanning temporal window, chosen for the Fourier transform (see Methods). In Fig. 3 (a) we present time-resolved profiles of the Brillouin frequency for the $H_2O$ ice X at 84 GPa, obtained by displacing the sample relative to the pump and probe co-focused laser beams in a lateral direction, i.e., parallel to the ice/diamond interface, by the steps of 1 μm. Both the TDBS signals in Fig. 3 (a) and the Brillouin frequency temporal profiles, which we call "images", in Fig. 3 (b) are shifted along the third axis for better visibility. Note that at 84 GPa the spatial in-depth distance of 1 μm corresponds to approximately 70 ps delay time.

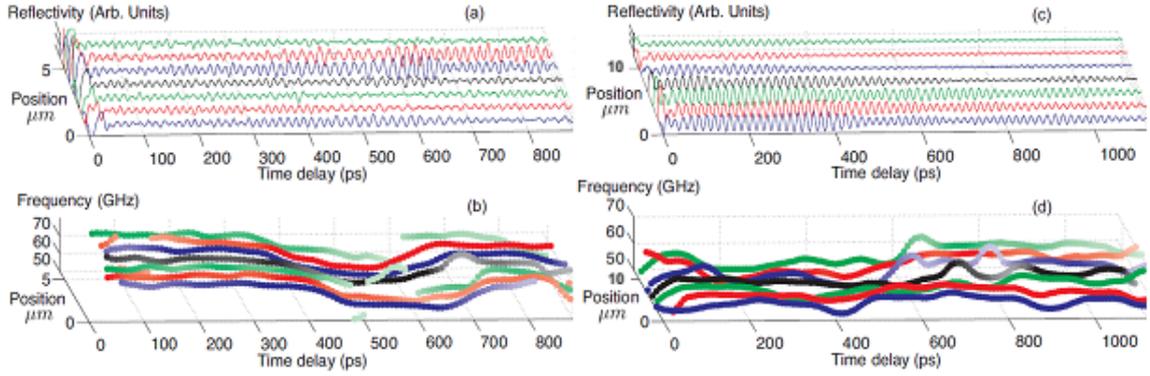

**Figure 3.** (a): TDBS signals in $H_2O$ ice at 84 GPa, obtained by displacing the sample relative to the co-focused pump and probe laser beams in lateral direction, i.e., parallel to the ice/diamond interface, by 1 μm steps. (b): Two-dimensional images of the Brillouin frequency magnitude obtained by processing the signals in (a). (c) TDBS signals in $H_2O$ ice at 57 GPa, obtained by displacing the sample relative to the co-focused pump and probe laser beams in lateral direction by 2 μm steps. (d): Two-dimensional images of the Brillouin frequency magnitude obtained by processing the signals in (c). The TDBS signals and temporal dependences are shifted along the third axis (Position, μm) for better visibility. The in-depth spatial resolution of the images is about 0.9 μm. In (b) and (c) signal amplitude is correlated with the symbols color: the darker the symbol - the higher the maximum amplitude of the Brillouin signal, and vice-versa.

The image in Fig. 3 (b) reveals a clear large-scale layering of the ice aggregate in the direction normal to the diamonds surfaces. The thickness of the layers in this large-scale texture is around 3 – 5 μm. In addition, smaller scale inhomogeneities, with the thickness about 1 μm controlled by the intentionally reduced spatial resolution, are observed at 84 GPa at some depth positions. The results of two-dimensional opto-acousto-optical imaging at 57 GPa presented in Fig. 3 (c) and (d) are obtained similarly to those in Fig. 3 (a) and (b) but using a larger step of 2 μm of lateral scanning. The image in Fig. 3 (d) reveals two regions, with much less pronounced than at 84 GPa in-depth layering, separated laterally. Visually the lateral separation in Fig. 3 (d) takes place in the middle of the laterally scanned region. The inhomogeneities with the spatial scale about 1 μm, controlled by the intentionally reduced spatial resolution, are also observed at 57 GPa at some depths. The results presented in Fig. 3 clearly demonstrate the ability of the TDBS technique applied here to reveal texturing of transparent polycrystalline aggregates such as our ice sample at the spatial scale exceeding 1 μm.

The estimated from the X-ray diffraction data characteristic dimension of the individual crystallites in the ice X aggregate at the considered pressures of 57 – 84 GPa is 0.5 μm (see Methods). In order to check the sensitivity of the TDBS technique to in-depth inhomogeneities at sub-micrometer scale, we processed the signals in the moving temporal window of 15 ps duration, approximately corresponding to a single period of the Brillouin oscillation at 84 GPa. The expected spatial resolution of about 0.2 μm is still controlled by the duration of the moving window. The results of the processing of the TDBS signals of Fig. 3 (a) in the delay time interval from 100 ps to 400 ps are presented in Fig. 4. Note that the temporal images are presented starting at the 100 ps delay time in order to avoid the region of pulse propagation in the vicinity of the Fe photo-generator where the results are commonly less precise because of non-perfect elimination of the high-amplitude thermo-reflectance contribution from the time-resolved reflectivity data.

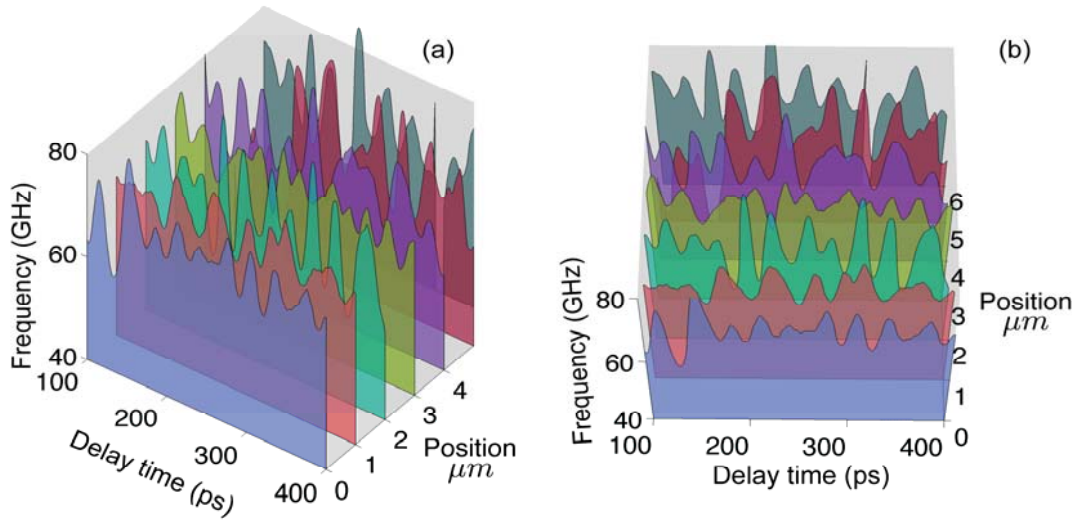

**Figure 4.** Two-dimensional depth profiles of the Brillouin frequency in $H_2O$ ice at 84 GPa, obtained by displacing the sample relative to the pump and probe co-focused laser beams in a lateral direction, i.e., parallel to the ice/diamond interface. The profiles in the figure are given in six particular positions of laser beams shifted with respect to each other in the same direction by 1 μm. The duration of the moving time-window for the Fourier transform is 15 ps, providing the in-depth spatial resolution of 0.2 μm. While the 3D representation in (a) makes clear the spatial ordere of the measurements, the view in (b) provides a better understanding of both in-depth and lateral structuring. The color code for the measured velocity profiles is the same in both images.

Figure 4 shows strong variations of the BS frequency (up to ±10 %) and, accordingly, of the longitudinal sound velocity by moving in-depth (corresponds to the delay-time axis). The characteristic spatial scale of these variations is 0.2-0.6 μm. We should bear in mind here a relatively large of ~ 10 μm$^2$ cross section of the opto-acoustically tested volume and a small size of grains of ~0.5 μm implies that the recorded BS signal (at any delay time ) is averaged in both lateral directions over about 40 individual grains. A large change of the BS frequency by moving in-depth to the next layer of grains (increase or decrease of the delay time by 20-30 ps) suggests a significant degree of crystallographic ordering in each of the layers but also a significant difference in the average crystallographic orientation of the adjacent ordered layers. This could be considered as a strong indication of the sample texturing parallel to the anvil culets and perpendicular to the direction of the sound pulse propagation. On the other hand, we do not recognise any pattern by shifting of our sample in a lateral direction (Fig. 4) even though the used step of 1 μm implies a partial overlapping of the examined volumes of the adjacent depth profiles of the BS frequency. This observation excludes the above discussed possibility of the sample texturing parallel to the diamond anvils, at least on the scale of 1 μm, in the lateral directions. The only possible model which we can suggest at the moment is that the sample body contains small mesoscopic groups of oriented crystallites. These groups could have the shape of discs or lenses. The depth/height of the oriented groups should be comparable with the size of one grain (0.3-0.6 μm) and the lateral dimension/diameter should not significantly exceed 1 μm. That means that such a grain contains between 4 and 10 ice crystallites with a preferential mutual ordering. Thus, the measurement with a small time-window of 15 ps could have revealed a some degree of texturing of our ice sample compressed to 84 GPa at sub-μm scale. However, the size of ordered/coherent regions is small and in lateral direction does not significantly extend over 1 μm.

Note, that characteristic direction of texturing, parallel to the diamond culets, revealed at large scale in Fig. 3 (b) and at short scale in Fig. 4, appears to be plausible if we consider the geometry of the sample and the uniaxial compression of the sample confined in the gasket.

It is worth noting here that the confirmation by the TDBS scattering technique of the sub-μm size of individual crystallites provides us opportunity to make a general statement that the averaging in our present experiments is not caused by the large number of crystallites scattering probe light in the tested volume but also to the strong diffraction of the scattered light on its way to the detector. Actually, for the probe wavelength of blue light of about 0.22 μm in ice X at 84 GPa, a part of the probe laser beam scattered in a grain with the diameter not exceeding 0.5 μm will diffract at distances shorter than 1 μm. We estimate the diffraction length by the ratio of the surface area of the scatterer or emitter to the wavelength. Thus the corrugations of the probe wave front caused by its scattering at the corrugated front of the coherent acoustic pulse propagating simultaneously in multiple crystallites is effectively smoothed by the diffraction at sub-μm scale. Thus, the detected signal provides information of the phase variation of the smoothed front of the scattered probe light and, consequently, on diffraction-averaged Brillouin frequency.

**2.2. On the role of the induced optical anisotropy**

In optically anisotropic grains the light experiences double or triple refraction and the optical rays, characterized by different refractive indexes, propagate after refraction with different velocities, which in addition depend on the direction of rays propagation relative to crystallographic axes.[48,49] However the ices VII and X of our interest here are cubic and, thus, optically isotropic. Consequently, the variations of the refractive index $n$ could be caused in our samples only by anisotropy induced by nonhydrostatic stress component, i.e., due to uniaxial loading.[50] Strong induced anisotropy could potentially cause splitting of the Brillouin frequency peak in TRBS experiments conducted with linearly polarized light.[51] The splitting effect in polarized probe light is replaced by broadening and shift of the Brillouin frequency line in our experiments conducted with circular polarized light. Note that even under uniaxial loading cubic crystals in general become biaxial[48] and exhibit triple refraction. An order-of-magnitude estimates demonstrate that induced optical anisotropy is negligible in our experiments. For example, in the experiments claimed to be conducted at 84 GPa, this pressure was measured in the centre of the DAC where the data were accumulated, while at the edge of the ice sample at the distance $R=D/2=45$ $\mu m$ from the centre the measured pressure was by $\Delta P = 4$ GPa lower. Thus uniaxial stress $t$ can be estimated from $t \approx H(\Delta P / R)$, e.g. Ref.[52] where the thickness $H$ of the sample is 13.5 $\mu m$, to be about $t \approx 1$ GPa. The characteristic value of the photo-elastic (acousto-optic) parameter $\partial n / \partial P$ can be estimated from the experimental curve $n = n(P)$ in Fig. 3 of Ref. 25 by $\partial n / \partial P \approx 10^{-3}$ 1/GPa at 84 GPa. Thus, an order-of-magnitude estimate for the possible difference in the refractive indexes of different optical rays in nonhydrostatically compressed $H_2O$ ice is $\Delta n \propto (\partial n / \partial P)t \approx 10^{-3}$. For the refractive index of $n \approx 1.8$, estimated from the data in Ref.[39] at $P=84$ GPa for the probe wavelength of 808 nm, this results in $\Delta n / n \propto 5 \cdot 10^{-4}$, indicating that the variations of the refraction index in Eq. (1) is largely negligible. However, in view of the complexity of the phenomena of stress-induced anisotropy and of the lack of data on the photoelastic tensor of $H_2O$ ice VII and X, we conducted additional test experiments with variable linear polarisation of the probe beam. In Fig. 5 we present three dependences on time of the Brillouin frequency, which are extracted from TDBS signals detected at 50 GPa in the same position of the sample but using instead of circularly polarized probe light linearly polarized light of three different polarizations obtained by polarization rotation in steps of 60 degrees. The results in Fig. 5 indicate that in our experimental system there are polarization-dependent effects. However, there could be different physical phenomena contributing to the variation of the Brillouin frequency with the polarization of probe light in Fig. 5. The experimentally observed shifts in Brillouin frequency could be attributed to the stress-induced optical anisotropy of cubic ice only under the conditions that only the polarization of the probe light incident on the surface of the ice is changing, when we rotate its polarization outside the DAC. Only in this case the experiments provide direct access to the refractive indexes of the different probe rays in cubic ice with stress-induced anisotropy, because the only variable parameter influencing the detected Brillouin frequency will be the distribution/repartition of probe light among different rays interacting with the acoustic pulse. However, in general, this ideal condition is difficult to achieve because, firstly, the position of light beam incident on the diamond, and consequently on the ice, could be slightly shifted when optical elements are used to rotate its polarization and, secondly, the position of the probe, the shape of the light focus and the polarization of the probe on the ice surface all could be influenced by polarization-dependent optical phenomena in diamond, caused by its natural optical anisotropy and important induced optical anisotropy due to strongly nonuniform stress generated in the uniaxially loaded anvils.[53-55] For example, the shift of the probe focus and modification of its shape on the surface of the ice could cause the shift of the detected Brillouin frequency even in case of negligible induced optical anisotropy of cubic ice if the photo-generated acoustic pulse propagates simultaneously across just two or more crystallites with different sound velocities along acoustic pulse propagation direction. In this case rotation of probe light polarization incident on the DAC could cause the redistribution of the probe light incident on the different parts of the acoustic beam cross section and to possible modification of the Brillouin frequency, because different cross sections of the acoustic beam are propagating with different velocities. It is worth mentioning, that stresses in the diamond anvils cause optical anisotropy of the anvils even close to their symmetry axis. They were reported to result, for example, in distortion/transformation of the initially linearly polarised laser light into elliptical.[56] Although currently we are not able to identify precisely the physical reasons of the polarization-dependent effect reported in Fig.5, we believe this experimental test clearly indicates the weakness of the effect. The relative changes of Brillouin frequency in Fig. 5 do not exceed 2 %. Thus the variations of the refractive index in Eq. (1) cannot be responsible for the 20% variations in the Brillouin frequency revealed in Figs. 2 and 4 and could be neglected when using Eq. (1) for the estimates of the acoustic velocity.

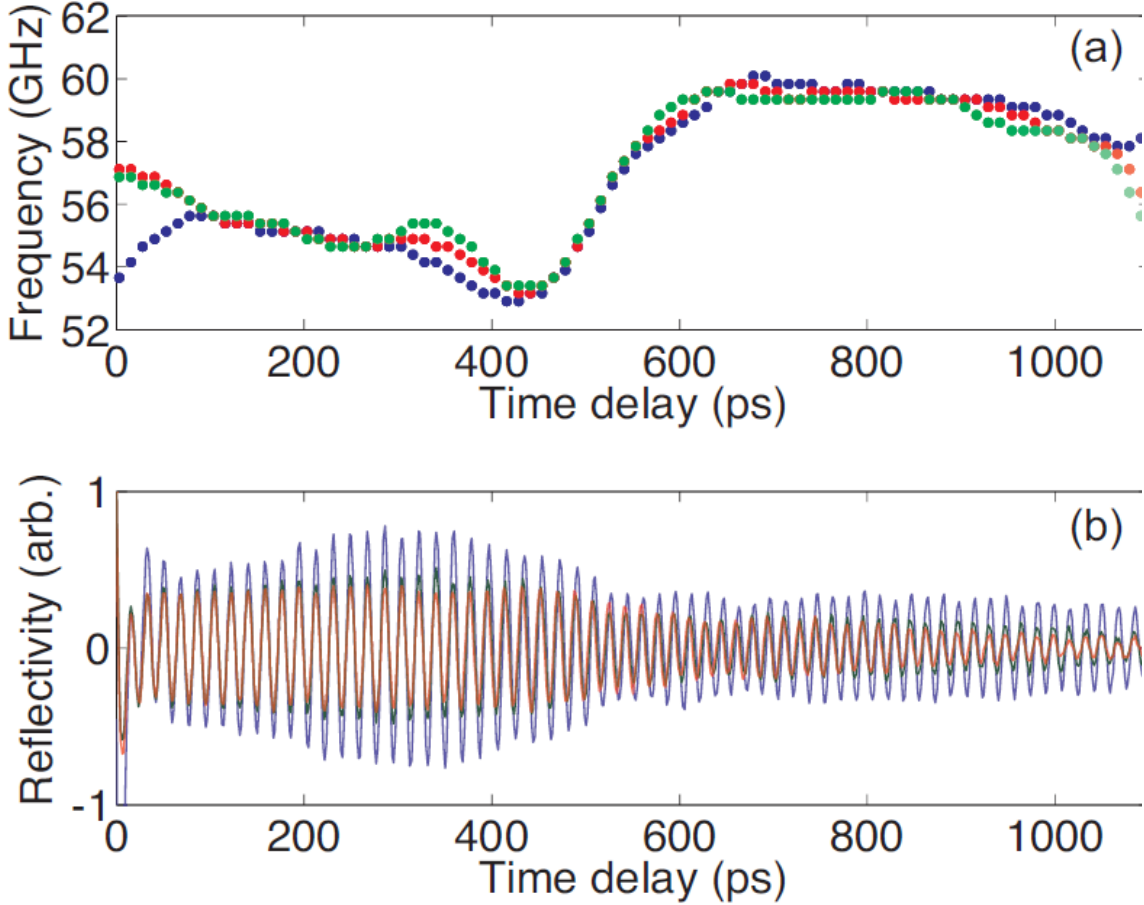

**Figure 5.** (a):Variations of the Brillouin frequency with time extracted from TDBS signals detected by linearly polarized probe light pulses of three different linear polarizations and presented in (b).

**2.3. On the role of the acoustic rays refraction**

The following considerations indicate that the dependence of the BS frequency on $\cos\theta$ can be neglected in Eq. (1) in comparison with the dependence on the sound velocity. When the acoustic wave is incident obliquely from an elastically isotropic medium (1) on its interface with an elastically isotropic medium (2) the direction of the transmitted wave shifts from the direction of the incident wave by a certain angle $\theta'$. If we denote by $\upsilon_{1,2}$ the sound velocities in the first and the second medium, respectively, then in the case $\upsilon_1 > \upsilon_2$ the maximum refraction takes place at grazing incidence ($|\theta'|_{max} = \pi/2 - \arcsin(\upsilon_2/\upsilon_1)$), while in the case $\upsilon_1 < \upsilon_2$ it takes place at critical angle of incidence $\theta_{cr} = \arcsin(\upsilon_1/\upsilon_2)$ ($|\theta'|_{max} = \pi/2 - \theta_{cr} = \pi/2 - \arcsin(\upsilon_1/\upsilon_2)$). It is worth noting that refraction angle relative to the initial propagation direction, accumulated through successive incidence at different angles on multiples interfaces, can't exceed the maximum refraction angle of single scattering. The maximum refraction angle obviously diminishes with diminishing difference between the acoustic velocities in the contacting materials. In the limiting case of small difference of sound velocities, denoting $\upsilon_1 = (\upsilon_1 + \upsilon_2)/2 + (\upsilon_1 - \upsilon_2)/2 \equiv \langle\upsilon\rangle + \Delta\upsilon$ and $\upsilon_2 = \langle\upsilon\rangle - \Delta\upsilon$, with $|\Delta\upsilon|/\langle\upsilon\rangle \ll 1$, we estimate $|\theta'|_{max} \approx |\Delta\upsilon|/\langle\upsilon\rangle$. This clear asymptotic result indicates that in polycrystalline samples with a weak elastic anisotropy of grains, i.e., with $|\Delta\upsilon|/\langle\upsilon\rangle \ll 1$, the contribution to relative variations of the Brillouin frequency $\Delta f_B/\langle f_B\rangle$ in Eq. (1), which is coming from the directional variations in the sound velocity and is proportional to $\Delta\upsilon/\langle\upsilon\rangle$, dominates over the contribution due to possible refraction of sound, which is proportional to $\Delta(\cos\theta)/\langle\cos\theta\rangle \le 1 - \cos(|\theta'|_{max}) \approx (\Delta\upsilon/\langle\upsilon\rangle)^2 \ll |\Delta\upsilon|/\langle\upsilon\rangle$. In these estimates, for the values of $\upsilon_{1,2}$ we should use the values of the maximum and the minimum values of the direction-dependent sound velocity in the microcrystals. From the experimental data on the elastic moduli of $H_2O$ ice at 40 GPa presented in Ref.[38] we estimate that the maximum longitudinal velocity (in [111] direction) is about 20% higher than the minimum longitudinal velocity (in [100] direction). This provides $|\Delta\upsilon|/\langle\upsilon\rangle \le 0.1$ and indicates that the contribution of variations in $\theta$ to the variations in the Brillouin frequency in Eq. (1) is at the level below 1%. Although in accordance with Ref.[38] the anisotropy increases with increasing pressure, we believe that even in our experiments, which are conducted at higher pressures, the variations in the Brillouin frequency from grain to grain are largely dominated by possible changes in the sound propagation velocity and not in the sound propagation direction. In fact in our experimental results in Fig. 3 revealing large-scale texturing both at 57 GPa and 84

GPa the maximum changes of the Brillouin frequency relative to the average level are about 10% ( $|\Delta f_B|/\langle f_B \rangle \leq 0.1$ ) and then $\Delta(\cos\theta)/\langle\cos\theta\rangle \leq 0.01$. Even when superposing the short-scale texturing revealed in Fig. 4 with the large scale texturing in Fig. 3 (b) the maximum changes of the Brillouin frequency relative to the average level are about 20% ( $|\Delta f_B|/\langle f_B \rangle \leq 0.2$, $\Delta(\cos\theta)/\langle\cos\theta\rangle \leq 0.04$ ). Consequently, in accordance with above presented estimates, the dependence on $\cos\theta$ in Eq. (1) is largely dominated by the dependence on the acoustic velocity.

In our opinion, there is also an additional factor, specific to micro-crystalline samples, which could significantly diminish the maximum effective angles that should be used in the estimates based on Eq. (1). We expect that propagation of the coherent acoustic beam quasi-collinear to the DAC axis is supported by the diffraction of the acoustic waves inside the beam. In fact, when acoustic pulse crosses a layer of differently oriented crystallites, distributed laterally in the pulse cross-section, the phase front of the acoustic pulse has a tendency to corrugation, because of the difference in the magnitudes and the directions of the sound velocity in different crystallites. However, this tendency is strongly suppressed by the diffraction phenomena smoothing the amplitude and phase differences between the different parts of the acoustic beam. The shortest 30 nm length of the acoustic pulse, realized in our experiments in the vicinity of Fe opto-acoustic generator, corresponds to the characteristic acoustic wave length of 0.2 μm (in ice X at 84 GPa). Thus the lateral corrugations of the acoustic pulse, which could accumulate in its refraction inside the crystallites with the dimensions below 0.5 μm, are continuously washed out by the diffraction at distances below 1.2 μm. Thus the diffraction inside the acoustic beam supports the propagation of all parts of the beam quasi-collinear to the DAC axis, while the diffraction of the total acoustic beam plays negligible role in our experiments. The diffraction length of the acoustic beam of 5 μm FWHM, equal to the short axis of the elliptical pump laser focus, exceeds the thickness of ours sample even for the waves of 2 μm length.

### 2.4. On the amplitude of the TDBS signal

The following considerations indicate the dominant role of the photo-elastic and elastic anisotropy of cubic ice in the experimentally observed variations of the TDBS signal amplitude by propagation through the sample. In principle, the non-monotonous changes in the amplitude of the oscillations revealed in TDBS (Fig. 1 (c)) could be caused by beatings/interference between two or more oscillations at different frequencies. Simultaneous detection of several frequencies by TDBS would be expected if, for example, the elastically isotropic opto-acoustic generator, Fe-foil in our case, launches in elastically anisotropic $H_2O$ ice not only the quasi-longitudinal acoustic pulse, which currently dominates in all our signals, but also one or two quasi-traverse acoustic pulses, propagating at velocities importantly slower than the quasi-longitudinal one and resulting in TDBS signals at very different frequencies. The generation of quasi-shear acoustic pulses by mode conversion of the longitudinal acoustic pulses at the interface between isotropic light absorbing photo-acoustic generator and anisotropic transparent elastic media was earlier reported,[57] and the beatings in the TDBS signals, caused by the interference of the Brillouin oscillations induced by quasi-longitudinal and quasi-shear waves, were observed.[57-61] Our experiments currently indicate that either the mode conversion of the longitudinal pulse into quasi-shear pulses at the Fe/ice interface and the mode conversion of quasi-longitudinal pulse into quasi-shear pulses at the interfaces between differently oriented crystallites are inefficient or the photo-elastic detection of quasi-shear waves is inefficient, or both. Several frequencies can be potentially found in TDBS signals in optically anisotropic media because of splitting of the probe laser beam in such media into two or three independent beams, which are characterized by different values of the refractive indexes.[51] However, cubic $H_2O$ ice is optically isotropic and our additional testing measurements have demonstrated that optical anisotropy, induced by possible deviation of mechanical loading of crystallites in DAC from the hydrostatic conditions, is negligibly small (see Fig. 5). It is also worth mentioning here that the period of possible beatings in TDBS signal amplitude, when two oscillations at different frequencies, $f_1$ and $f_2$, are contributing to it, is equal to $1/|f_1-f_2|$. Thus, even if we estimate the difference between frequencies corresponding to ordinary and extraordinary light, $|f_e-f_o|$, by double frequency shift detected in Fig.5, i.e., about 2 GHz, the expected beating period will be 500 ps, i.e., much longer than the time interval where birefringence is observed in this Figure. This is an additional argument in favor of the conclusion that the observed non-monotonous variations of TDBS signal amplitude are not caused by optical anisotropy.

Finally, the detection of several frequencies could be expected in polycrystalline samples if the photo-generated acoustic pulse and the probe light propagate simultaneously across several differently oriented crystallites. This could be possible not only if the diameters of the acoustic and probe laser beams are importantly larger than the characteristic dimension of the grains in lateral direction, i.e., in the plane parallel to the surface of the photo-generator, but also in the case when they are narrower than the individual grains but propagate along the interface of two grains, which is inclined relative to sound propagation direction, or they propagate along the interface separating larger groups of similarly oriented crystallites. In our experiments reported here the diameters of the acoustic and the probe beams are exceeding the estimated dimensions of the individual crystallites, so the first and the third of the mentioned situations seem to be more relevant. However, in both these cases the situations when the cross section areas of the acoustic pulse in two differently oriented domains of ice are close in magnitude, leading to two oscillations of comparable amplitudes and to possibility of important beatings, are rear. In fact, in the experimental data presented in Fig. 3 the signal processing has not revealed two frequency peaks of comparable amplitudes along the complete paths of the photo-generated acoustic pulses. The above discussion leads to the conclusion that non-monotonous variations of the TDBS signal amplitude in time is not the manifestation of the beatings/interference among different frequency components of the signal but is rather the second, additional to Brillouin frequency variation in time, direct manifestation of disorientation of grains or group of grains in polycrystalline $H_2O$ aggregate. Actually, cubic crystallites of ice being optically isotropic are anisotropic photo-elastically (acousto-optically).[49,62] The magnitude of the effective photo-elastic constant, which couples collinearly propagating sound and light, depends on the direction of their propagation inside the individual crystallite. Then the amplitude of the time-domain Brillouin oscillation is expected to be different in the crystallites differently oriented relative to the sample normal. Since the amplitude $A$ of the Brillouin oscillation is directly proportional to the effective photoelastic constant and inverse proportional to the square root of the acoustic velocity, i.e,, inverse proportional to the root of the forth power from the effective elastic modulus ( $A \propto pn^{-1}(\upsilon\rho)^{-1/2}$, where $p$ and $\rho$ denote the photo-elastic constant and density, respectively),[32,63] it could be expected that the photo-elastic anisotropy provides the dominant contribution in comparison with elastic anisotropy to the non-monotonous variations of the amplitude of the Brillouin oscillations in polycrystalline aggregates but does not obscure the frequency variations due to the elastic anisotropy of the examined sample. The preliminary analysis of the amplitude variations, accomplished under the same conditions as that of the frequency analysis presented in Fig. 4, demonstrates that they are highly correlated in time with frequency variations, confirming that the physical origin of both is in the orientational texture in our sample. The preliminary analysis also indicates that the magnitude of the amplitude variations significantly exceeds what could have been expected due to the variations of the acoustic velocity only, i.e., $\Delta A/A \propto -(1/2)(\Delta \upsilon/\upsilon) \propto -(1/2)(\Delta f/f)$. This indicates an important role of acousto-optic (photo-elastic) anisotropy.

**2.5. On the duration of the photo-generated acoustic pulse and ultimate in-depth spatial resolution of TDBS**

In depth-profiling of the spatially inhomogeneous media the in-depth spatial resolution of TRBS is known to be limited either by the spatial length of the coherent acoustic pulse or by restrictions introduced by specific signal processing techniques.[32,33] In our experiments the coherent picosecond acoustic pulse is predominantly generated by the thermo-elastic mechanism in the iron foil, because of the negligible energy transport from iron into ice caused by weak thermal diffusivity of ice in comparison with that of iron.[64] As the duration of the picosecond blue pump laser pulse (1.9 ps at FWHM) exceeds the characteristic time of the longitudinal acoustic wave propagation across the region of the absorbed optical energy release in iron, the duration of the photo-generated acoustic pulse is controlled by the duration of the laser pulse.[20] Note that the depth of lattice heating in Fe does not importantly broadened relative to the depth of the blue light penetration, because of the relatively strong coupling of the electrons and the phonons in Fe. Even for the overestimated depth of the energy release zone, 15 nm at FWHM, the time of sound propagation across it with the velocity of $8.4 \cdot 10^3$ m/s Ref.[65] is shorter than the duration of the laser pulse. So the length of the coherent acoustic pulse launched in ice near its interface with the Fe foil is estimated to be of about 30 nm in $H_2O$ ice at pressures of the experiment. This provides the best possible in-depth spatial resolution unless the amplitude of the launched pulse is so strong that the nonlinear acoustic effects could lead to weak shock front formation. In the latter case, which is actually not realized in our experiments, the spatial resolution could be controlled by the width of the shock front, which could be shorter than the width of the initially launched acoustic pulse.[66,67] In our linear case the situation is, in a certain sense, opposite to the case of the nonlinear acoustics. Due to the preferential absorption and scattering of high frequency components, the initially launched coherent acoustic pulse is continuously broadening along its propagation path. However, the abrupt variations in the Brillouin oscillations amplitude and frequency observed experimentally (Figs. 1 (c) and 2 (a)) indicate that the acoustic pulse is shorter than the Brillouin period even when arriving on the ice/diamond interface. Thus the duration of the acoustic pulses does not influence the spatial resolution of the images obtained in the current Report with the moving windows, which are equal to or larger than the Brillouin period.

**2.6. On the measurements of aggregate and envelope sound velocities**

We applied optical interferometry to measure the product $H \cdot n$ of the ice layer thickness between Fe/ice and ice/diamond interfaces $H$ (see Fig. 1 (b)) and the optical refractive index $n$ of $H_2O$ ice at the probe beam wavelength. Then using the values of $n$ extrapolated from the published data[39] we estimated $H$. We measured the propagation times of the acoustic pulse between Fe/ice and ice/diamond interface at each pressure in three different positions of the sample via determination of the time moment when the abrupt change in the Brillouin oscillation amplitude takes place, estimated three averaged velocities and found their average value. Thus estimated sound velocities at 84 GPa and 56 GPa are 15500 m/s and 13500 m/s, respectively, in a very good agreement with the values of aggregate sound velocities measured by classic BS.[36] We also applied the so-called envelop method[36,38] to find $C_{11}$ modulus in our samples. For this we found the minimum value of the Brillouin frequency from the data detected in three different points of the samples, and determined the minimal longitudinal wave velocity, using $n$ extrapolated from Ref.[39]. The minimum velocity in cubic crystallites is along the [100] direction and depends only on $C_{11}$ and the density of ice. Taking the values of densities reported in Ref.[44] we found the values of the elastic modulus $C_{11}$ in a good agreement with the values extracted by classical BS.[36] A detailed report on the sound velocities of ice VII and X as a function of pressure will be presented in a separate paper. The main topic of our present Report is demonstration of the imaging capabilities of the TDBS technique applied to optically transparent polycrystalline aggregates.

## 3. Discussion

In this work we were able to approach a high in-depth spatial resolution of about 0.2 μm in ice sample compressed in a DAC to 84 GPa. This value is significantly below the estimated average crystallite size in this sample. Apparently, this result can be achieved for any other transparent material, crystalline or amorphous, compressed in a DAC to similar or higher pressures. No other technique can provide today such a detailed visualisation of a polycrystalline aggregate microstructure of a chemically homogeneous material compressed in a DAC, especially along the axial direction. Even the simplest signal processing, applied in the present work for processing of the TDBS signals, provides access to characterization both of the in-depth dimensions of the micro-crystallites and of the texturing of the polycrystalline ice aggregates at different spatial scales from 0.2 μm, controlled by signal processing, up to about 10 μm, controlled by the total thickness of the ice layer. It is worth mentioning here, that TDBS technique was applied earlier to the diagnostic of the growth of the sub-micrometer thick ice layer from the vapour phase at about 110 °K.[68] However, although the ice layer was supposed to be polycrystalline, no attempts to reveal the signatures of its polycrystallinity or texturing were undertaken.

From the experimental geometry used in the present work follows that even when the shortest moving time-window for the Fourier transform of 15 ps is used, the collected TDBS signal and the derived frequency/sound velocity is an average over about 40 grains. This is because the grain size was estimated to be about 0.5 μm, the lateral surface area tested by overlapping probe laser beam and coherent acoustic beam is about 10 μm$^2$ and the wavelength of the probing blue laser light in ice at Mbar pressures is approximately 0.2 μm. Thus, it is difficult to access in the present sample and experiment geometry, at any position in the sample, the extreme sound velocity values, corresponding to the specific directions in a single crystal of ice X. The highest sound velocity observed in the present work by scanning through the sample will be lower than the maximal possible sound velocity in a single crystal ice X (along the <111> direction of a cubic crystal). Similarly, the lowest sound velocity detected here represents the upper bound for the lowest possible sound velocity in a single crystal ice X (along the <100> direction of a cubic crystal). .

One of the ways to approach the limiting values would be to use coarse powder or single crystals as starting samples. The grains could still break on pressure increase due to nonhydrostatic loading or after phase transitions but the fragments could still remain big enough to deliver to the detector the TDBS signals from the individual grains. In the particular case of water, single crystals of ice VI and VII can be easily obtained by a slow compression at about 1 GPa when crossing the melting curve e.g.[35,69]. Another way involves a significant improvement of the lateral resolution of the experimental set-up below 0.5 μm by application of advanced focusing methods. Combined with the already available high in-depth resolution, the improved lateral resolution would make possible determination of the limit values of sound velocities as well as a 3D-visualization of transparent polycrystalline aggregate microstructure (form grain to grain) at Megabar pressures.

The time-domain BS technique could have other, i.e., additional to improved spatial resolution, advantages in comparison with the classic frequency-domain BS technique. The applicability of classic BS is importantly reduced and even could be impossible in the case when the Brillouin spectral lines of the sample overlap with the Brillouin lines from the diamond anvils[24], which are inevitably simultaneously detected for samples compressed in a DAC. This problem does not exist for the time-domain BS technique because the scattering of light by coherent acoustic phonons takes place in the sample only, well before the photo-generated acoustic pulse reaches the sample/diamond interface. It is also an advantage of TDBS that slowly varying thermo-reflectance signals, caused by pulsed laser heating of optoacoustic transducer, which obscure the Brillouin oscillation just after the photo-generation of the coherent acoustic pulse, are progressively diminishing with delay time, while the Rayleigh scattering line in the classic BS, which obscure the Brillouin scattering lines, is always present.

4. **Conclusions**

The developed here TDBS – based imaging technique provides for each crystallite (or a group of crystallites) in a chemically homogeneous transparent aggregate usable information on its orientation (if the material is elastically anisotropic) as well as on the value of the elastic modulus along the direction of the sound propagation. This extends the basis for a successful application of highly developed micromechanical models of visco-plastic deformation of solids at Mbar pressure. On long term, such experiments extended to Earth's minerals and high or low temperatures would insure a significant progress in understanding of convection of the Earth's mantle and thus evolution of this and other planets.

The achieved two-dimensional imaging of the polycrystalline aggregate in-depth and in one of the lateral directions indicates the feasibility of three-dimensional imaging of transparent samples compressed in a DAC with tens of nanometers in-depth resolution and lateral spatial resolution controlled by the pump and the probe laser pulses focusing. In perspective, an improved signal processing of such TDBS data should provide opportunity to follow evolution of several Brillouin frequencies in time domain, revealing simultaneous propagation of the coherent acoustic pulse across several mutually disoriented crystallites. In the future, TRBS experiments, conducted both with longitudinal and shear[70-72] coherent acoustic pulses at several different angles of probe light incidence[32,46,47] could give access to the determination of the spatial positions, dimensions and orientation, i.e., morphological and orientational texture, of the optical refractive index and of all elastic moduli of the individual grains inside polycrystalline transparent aggregates compressed to Megabar pressures.

**Methods**

**Diamond anvil cell and the samples**

High pressures up to 84 GPa were generated via squeezing of samples between bevelled diamond anvils of Boehler-Almax design having the culet size of 300 μm mounted in a Boehler-Almax

Plate DAC.[73] A hole in the centre of a pre-indented stainless steel gasket represented the sample volume filled with ice and a thin iron foil in contact with one of the anvils and the sample served as the opto-acoustic generator for launching coherent acoustic pulses in the ice. A zoomed schematics of the sample arrangement in the DAC is presented in the insert in Fig. 1 (b). The sample dimensions in the experiments conducted at 57 GPa and 84 GPa were, respectively, 103 μm and 90 μm in diameter $D$ and 14.4 μm and 13.5μm in thickness $H$. The diameter $d$ of the iron opto-acoustic generators were 66 and 40 μm, respectively and their thicknesses about 2 μm at ambient pressure, in both experiments. Pressure was determined from the wavelength of the $R_1$ fluorescence line of ruby grains distributed through the ice sample which red shift with pressure was calibrated earlier.[74]

**Pump and probe optical schema**

The experiments on ice compressed in the DAC were performed using a typical pump/probe configuration for transient reflectivity optical measurements (Fig. 1 (a)) involving a pulsed Ti:Sapphire laser with the following characteristics: 2 W average power, 808 nm wavelength, 2.7 ps FWHM duration of the laser pulses at the repetition rate of 80 MHz. This radiation was divided by a polarizing cube in the pump- and probe beams. The pump laser beam was modulated acousto-optically at frequency 161.1 kHz for the subsequent realization of the synchronous detection of the probe laser radiation scattered by the sample. Then, it was frequency-doubled by 1 cm-long BBO non-linear crystal to obtain 25 mW of 404 nm wavelength light pulses of 1.9 ps duration at FWHM for the generation of coherent acoustic pulses in Fe near its interface with $H_2O$ ice. Computer-controlled optical delay line of two-passage configuration allowed introducing the delays of the probe laser pulses relative to the pump laser pulses in the interval 0-8 ns. The time-delayed probe laser pulses, for time-resolved detection of the BS induced by the coherent acoustic pulse, were at the fundamental wave length 808 nm of the laser. The photo-acoustic signals have been obtained with the steps in time of 1.5 ps. Both pump and probe radiation were focused by a 50X-objective lens (numerical aperture 0.5, working distance 10.5 mm on the DAC from the same side and the scattered probe radiation was also gathered from the same side. The system of optical imaging including a web-camera and the source of white light has been installed for the visualization of the laser spots and the surface of the sample in DAC. The pump laser beam was focused on the Fe/ice interface into the elliptical spot, which dimensions, 9 μm in the horizontal and 5 μm in the vertical directions, are determined at FWHM of the intensity of the image obtained by the camera. The probe laser pulse was focused into the circular spot. The pump spot was scanned relative to the probe in vertical lateral direction along the Fe/ice interface perpendicular to the long axis of the ellipse. The scan was achieved using a variation of an angle $\Delta\theta$ of incidence of pump beam, provided by the rotation of a dielectric mirror installed on the computer controlled support (M2 in Fig. 1 (a)). The displacement $\Delta x$ of the pump beam on the surface of Fe is the following: $\Delta x = [f - h(1 - 1/n)]\Delta\theta$, where $f$=4 mm is the focusing distance of the objective, $h$=2.5 mm and $n$=2.4 are the thickness and the refractive index of diamond, respectively. The position of the delay line was fixed at the maximum of amplitude of the transient thermo-reflectance signal, which corresponds to the coincidence in time of pulses of pump and probe beams. Then, the pump beam was scanned in vertical direction to obtain the amplitude of the transient thermo-reflectance signal as a function of the position. This function, showing the correlation of the pump and probe beams in the vertical direction, has provided the width of 3 μm at FWHM. This indicates a 4 μm radius of the probe laser focus and 4.5 μm FWHM of the correlation function of the two beams in the direction of the long (horizontal) axis of the pump laser elliptical focus. In addition, DAC was mounted on a motorized linear stage (M1 in Fig. 1(a)), allowing displacement of laser spots in horizontal lateral direction on the surface of the opto-acoustic generator with precision of 0.1 μm. The filter (F in Fig. 1 (a)) was introduced before the photo-detector to avoid its illumination by pump radiation scattered from the sample.

**Processing of TDBS signals**

The panels (b) and (d) in Fig. 3 represent the dominant frequency of the reflectivity signals as a function of time. The dominant frequency values at each time are obtained from a spectrogram analysis of the temporal signals with a Hanning weighted window of 122ps (128 points at 0.9595 THz sampling frequency). For each central time of this sliding analysis window, every ~8ps, the spectral component with the maximum amplitude is extracted. The frequency determination precision is improved by interpolating the spectrum with a spline function, which provides frequency steps of 0.2 GHz. The maximum amplitude is used for the color scale of the plotted symbols. Thus, the darker the symbol is, the larger is the maximum amplitude of the Brillouin signal, and vice-versa.

**Estimates of the crystallite dimensions from X-ray diffraction data**

We have collected two dimensional XRD patterns of the $H_2O$-iceVII (and X) samples compressed in a DAC using a synchrotron radiation of the beam-line P2.02 (Petra III, HASYLAB, DESY).[75] The patterns were collected from sample areas of about 15x15 μm$^2$, the sample thickness was also about 15 μm. The samples were rotated by about 4 degrees around the vertical axis in order to increase the number of grains in the Bragg diffraction condition. These measurement conditions led to a smooth without any gaps distribution of intensity along diffraction rings for all observed *hkl* reflexes, which suggests a relatively large number of randomly oriented crystallites covering the full solid angle 4π. In order to estimate the lower limit of the grains number needed for this, we need to know the solid angle covered by a single grain under the present experimental conditions. To find it we used for the divergence of the beam, which focussed the monochromatic synchrotron beam on our sample compressed in a DAC, the value of compound reflective lenses (CRL)[76] estimated to be 0.65 mrad.[77] Then the effective divergence is 1.2 mrad due to the fact that the spot on the 2D detector is almost 2 times larger than the pixel size of 0.2 mm. Combining the effective divergence with the angle of rotation around the vertical axis we obtained the solid angle covered by scattering from a single grain to be 8.5·10$^{-5}$. Taking into account cubic symmetry of crystal structure of ice VII and ice X we estimate the number of crystallites needed to cover the 4π solid angle by uniform XRD rings at about 25000. Accounting for the sample volume illuminated with the X-ray beam we estimate an average size of crystallites in our samples at 57-84 GPa to be ~ 0.5 μm. This value is close to the dimensions of regions with similar magnitude of the Brillouin frequency we revealed by processing the TDBS data in the moving time window of 15 ps (Fig. 4).

**Acknowledgements**

This research was conducted in the frame of the project "LUDACISM" supported by the program ANR BLANC 2011. We would like to thank Guy Dirras for fruitful discussions on texture and W. Morgenroth for assistance in using the Extreme Conditions Beamline (P02.2) of the light source PETRA III at DESY, a member of the Helmholtz Association (HGF). X-ray measurements were partially funded the European Community's Seventh Framework Programme (FP7/2007-2013) under grant agreement n° 312284.